\newcommand{\be}{\begin{equation}}
\newcommand{\ee}{\end{equation}}
\newcommand{\bea}{\begin{eqnarray}}
\newcommand{\eea}{\end{eqnarray}}

\newcommand{\lab}[1]{\label{#1}}
\newcommand{\bi}[1]{\bibitem{#1}}
\newcommand{\da}{\,da}

\newcommand{\half}{\frac{1}{2}}

\newcommand{\smhalf}{{\textstyle \frac{1}{2}}}
\newcommand{\smthtw}{{\textstyle \frac{3}{2}}}

\newcommand{\smfrac}[2]{{\textstyle \frac{#1}{#2}}}

\newcommand{\smsqrt}[1]{{\textstyle \sqrt{#1}}}
\newcommand{\aleq}{\mathrel{\rlap{\lower4pt\hbox{\hskip1pt$\sim$}}
                   \raise1pt\hbox{$<$}}}
\newcommand{\ageq}{\mathrel{\rlap{\lower4pt\hbox{\hskip1pt$\sim$}}
                   \raise1pt\hbox{$>$}}}

\documentstyle[aps,manuscript,epsf]{revtex}

\begin{document}
\title{Remark on approximation in the calculation of the primordial 
spectrum generated during inflation}
\author{D.H. Huang~$^{1}$,  W.B. Lin~$^{2}$,   X.M. Zhang~$^{3,2}$}
\address{
~\\$^1$
Deptartment of Physics, Peking University, Beijing 100871, P.R. China;
~\\$^2$
Institue of High Energy Physics, Chinese Academy of Science, P.O. Box 918-4, 
Beijing 100039, P.R. China;
~\\$^3$
CCAST (World Lab), PO Box 8370, Beijing 100080, P. R. China
}
\maketitle
~~\\
~~\\
\begin{abstract}
{We re-examine approximations in the analytical calculation of the primordial
 spectrum of 
cosmological perturbation produced during inflation. Taking two inflation 
models (chaotic inflation and natural inflation)
as examples, we numerically verify the accuracy of these approximations.} 
\end{abstract}


The inflationary prediction on the primordial power spectrum of the 
cosmological
perturbation is testable by the observations of the large-scale 
structures\cite{rep},
therefore it is important to make the prediction more precise. In 
this brief report,
we re-examine the analytical calculation of the primordial spectrum of 
cosmological perturbation produced during inflation and point out 
a subtlety 
in the approximations.

To begin with, we review briefly the derivation of the power spectrum 
using the standard slow rolling approximation (SRA) technique\cite{LS,SL}. 

The most general form of the metric for the background and the scalar metric 
perturbations 
can be written as
\cite{bardeen}
\be 
ds^2=a^2(\eta)\{(1+A)d\eta^2-2\partial_i B dx^id\eta
- [(1-2\Psi)\delta_{ij}
+ 2\partial_i\partial_j E]dx^idx^j\}~,
\ee
where $\eta\equiv \int dt/a(t) $ is conformal time.

The primordial spectrum of scalar perturbation is defined as 
\bea 
P_{S}(k) 
\equiv \frac{k^{3}}{2\pi^{2}}
\left|{\mathcal R}_k \right|^2
,                               \lab{szeta}
\eea
where ${\mathcal R}_k$ is the coefficient of the Fourier transformation 
of the intrinsic curvature perturbation of comoving hypersurfaces $\mathcal R$ 
which during inflation has the form\cite{SL,lidsey}
\be
{\mathcal R} \equiv -\Psi - \frac{H}{\dot \phi}\delta \phi~,
\lab{zeta}
\ee
where $\delta \phi$ represents the fluctuation of the inflaton
 field $\phi$ and
$H=\dot{a}/a$ is the Hubble constant. 

Note that $P_S(k)$ in Eq.(\ref{szeta}) by definition is constant, however,
as we can see later the power spectrum during the period of inflation 
evolves in time and approaches a constant only when $k/aH\rightarrow 0$.
For the convenience of our discussion we introduce $P_S(k,a)$, the power
spectrum of a given $k$ mode as a function of time, to distinguish from 
its asymptotic constant value 
$P_{S}(k) \equiv P_S(k,a)|_{k/aH\rightarrow 0}$.

To calculate $P_{S}(k,a)$, 
one usually defines $u_k \equiv -z {\mathcal R}_k$ 
where $z\equiv a\dot{\phi}/H$. 
And $u_k$ satisfies the following equation of motion\cite{MFB} 
\be
u_{k}''+\left(k^{2}-\frac{z''}{z}\right)u_{k}=0 ,             \lab{ueom}
\ee
where a prime denotes the derivative with respect to conformal time $\eta$.
 
According to the flat spacetime field theory (well inside the horizon), 
one has  
\be
u_{k} \rightarrow \smfrac{1}{\sqrt{2k}}e^{-ik\eta} \;\;\;\;
{\rm as}\;\;\; k/aH \rightarrow \infty .                            
\lab{ulim} \ee
On the other hand, the solution of Eq.(\ref{ueom}) in the limit of 
$k/aH \rightarrow 0$ is 
\be
u_k \propto z~. 	  \lab{ulim1}
\ee  
Eq.(\ref{ulim1}) indicates that $P_{S}(k,a)$ becomes approximatly constant for 
mode $k$ to be
well outside the Hubble radius, which can also be seen directly from the 
equation of motion of ${\mathcal R}_k$:
\be
\dot{\mathcal R}_k=\frac{2}{3}(1+w)^{-1}H\left(\frac{k}{aH}\right)^2 \Psi_k~,
\ee
here, $w=p/\rho$ is the ratio of the pressure to the density.

To describe the period of slow-rolling of inflation, one defines two 
parameters $\epsilon$ and $\delta$ \bea
\nonumber \epsilon &\equiv&\frac{-\dot{H}}{H^2} \ll 1~,\\
\delta &\equiv& \frac{\ddot{\phi}}{H\dot{\phi}} \ll 1~,
\eea
in term of which $z^{\prime\prime}/z$ can be expressed as\cite{SL}
\be
\frac{z''}{z} = 2a^{2}H^{2}\left(1+\smthtw\delta+\epsilon
+\smhalf\delta^{2}+\smhalf\epsilon\delta
+\smhalf\smfrac{1}{H}\dot{\epsilon}
+\smhalf\smfrac{1}{H}\dot{\delta}\right)~,                      \lab{zed}
\ee
and the conformal time 
\be
\eta = \int\frac{dt}{a} = \int\frac{da}{a^{2}H} =
\frac{-1}{aH}+\int\frac{\epsilon\da}{a^{2}H}~.                  \lab{tau}
\ee

Assuming $\epsilon$ and $\delta$ to be constant during the period of slow 
rolling, Eq.(\ref{ueom}) can be reduced to\cite{SL}:
\be
u_{k}''+\left(k^{2}-\frac{\nu^2-\frac{1}{4}}{\eta^2}\right)u_{k}=0~,           
  \lab{ueom1}
\ee
where $\nu = (1+\delta+\epsilon)/(1-\epsilon)+1/2$ and 
\be
\eta = \frac{-1}{aH}\left(\frac{1}{1-\epsilon}\right)~.
\lab{tauc}
\ee
We emphasize that to obtain Eqs.(\ref{ueom1}) and (\ref{tauc}),
one has neglected the terms proportional to 
$\dot \epsilon$ or 
$\dot \delta$ in Eq.(\ref{zed}) 
and taken $\epsilon$ in Eq.(\ref{tau}) as a constant.

The solution of
Eq.(\ref{ueom1}) for $u_k$ is 
\be
u_{k} =  \smhalf\smsqrt{\pi}e^{i(\nu+\half)\frac{\pi}{2}}
(-\eta)^{\half}H^{(1)}_{\nu}(-k\eta)~.                        \lab{us} 
\ee
In the limit of $k/aH \rightarrow 0$, $u_{k}$ becomes
\be
  u_k \rightarrow  e^{i(\nu-\half)\frac{\pi}{2}}2^{\nu-\frac{3}{2}}
\frac{\Gamma(\nu)}{\Gamma(\frac{3}{2})}\frac{(-k\eta)^{\half-\nu}}{\sqrt{2k}}
 ~. 
\ee                                                            \lab{uslim}

Substituting $\eta$ of Eq.(12) into Eq.(14), one has the power 
spectrum in the slow rolling approximation, 
\bea
P_{S}^{(SRA)}(k,a) = 
\frac{2^{2 \nu-3}}{4\pi^2}\frac{\Gamma^2(\nu)}{\Gamma^2(\frac{3}{2})}(1-\epsilon)^{2 \nu-1}
\frac{H^4}{ |\dot{\phi}|^2}\left( \frac{k}{aH} \right)^{3-2\nu}~, 
~~~~{\rm as}~~~~\frac{k}{aH}\rightarrow 0~.
\lab{prc}
\eea
To obtain the final value of power spectrum,
what has been done in the literature is to evaluate the time 
dependent $P_{S}^{(SRA)}(k,a)$ in the equation above at $k=aH$ 
\cite{LS,SL}:
\bea
P_{S}^{(SRA)}(k) &\simeq & Q(k,a)|_{aH=k} \\
&=& \frac{2^{2 \nu-3}}{4\pi^2}\frac{\Gamma^2(\nu)}{\Gamma^2(\frac{3}{2})}(1-\epsilon)^{2 \nu-1}
\left.\frac{H^4}{|\dot{\phi}|^2}\right|_{aH=k}~,
\lab{hca}
\eea
where $Q(k,a)$ is defined as the right hand side (R.H.S.) of Eq.(\ref{prc}).

Note that $P_{S}^{(SRA)}(k,a)$ in
Eq.(\ref{prc}) is valid only in the limit of $k/aH \rightarrow 0$,
however, to obtain Eq.(\ref{hca}) one takes $ k/aH = 1$. One wonders if 
the value in Eq.(\ref{hca}) is a good approximation to 
the asymptotic value $P_{S}(k)$ in the limit of
$k/aH \rightarrow 0$?  Furthermore, $Q(k,a)$ is mathmetically divergent for
$\nu > \frac{3}{2}$
and zero for $\nu < \frac{3}{2}$ in the limit of $k/aH \rightarrow 0$,
so $Q(k,a)$ can only be a description of the true evolution of
$P_{S}(k,a)$
for some specific range of $\ln (aH/k)$.
Therefore it is questionable to have the
asymptotic value of the real spectrum obtained from Eq.(\ref{prc}).

In this brief report,
we numerically solve Eq.(\ref{ueom}) and study the
evolution of $Q(k,a)$ and $P_{S}(k,a)$ after horizon crossing.
The initial conditions of $u_k$ and $u'_k$ are given
by Eq.(\ref{us}) when the mode $k$ crosses out the Hubble radius.
In the numerical calculation of the amplitude of 
the primordial spectrum 
we take two models as examples: one is chaotic inflation model with 
a potential $V=m^2 \phi^2/2$, 
and the other is natural inflation model\cite{adams} 
$V=\Lambda^4[1+\cos(\phi/f)]$. 
We consider only the amplitude of the primodial spectrum for 
the COBE scale ($k_{COBE}\simeq 7.5 a_0H_0$).
The fluctuation mode corresponding to this scale crosses out the Hubble radius 
at the time e-folds $N\simeq 57$ before the end of inflation. 
This condition determines the value of inflaton field $\phi |_{aH=k_{COBE}}$.
Our numerical results on the 
evolution of $P_{S}(k_{COBE},a)$ and $Q(k_{COBE},a)$ 
are presented in 
Fig.1 and Fig.2.
In each of these figures, 
a dashed horizontal line marks the exact asymptotic 
value of the power spectrum.

Fig.1(a) and Fig.2(a) show the evolution of $P_{S}(k_{COBE},a)$ 
after the COBE scale crosses outside the horizon during inflation.
From these two figures one can see that 
when $\ln(aH/k_{COBE})\approx 3$, $P_{S}(k_{COBE},a)$ starts 
to be approximatly a constant.
In Fig.3 and Fig.4 we show
the relative error between $Q(k_{COBE},a)$ and the exact numerical result of 
$P_{S}(k_{COBE},a)$ when $\ln(aH/k_{COBE})\geq 3$.
For $3\leq \ln(aH/k_{COBE})\leq 6$ it is very small,
so $Q(k_{COBE},a)$ in this range of $\ln(aH/k_{COBE})$, 
which we denote as $Q(k_{COBE},a)|_{\dot{\mathcal R}\simeq 0}$, 
will be a good approximation to the real spectrum $P_S(k_{COBE},a)$ 
in the same range,
and then can be taken reasonablly as 
the asymptotic value of the spectrum in the limit of $k/aH\rightarrow 0$.

Fig.1(b) and Fig.2(b) show 
the evolution of $Q(k_{COBE},a)$ after crossing outside 
the horizon.
From these figures, we can see that the uncertainty in calculating
 $Q(k_{COBE},a)|_{\dot{\mathcal R}\simeq 0}$ by using the approximate formula
Eq.(\ref{hca})
is of order $1\%$. We conclude that
the usual analytical treatment in obtaining the primordial power spectrum is 
feasible with a good accuracy.

Before concluding this report, we should point out that
Nakamura and Stewart\cite{NS} have tried to tackle this issue 
by expanding analytically  the R.H.S. of Eq.(\ref{prc}) at $k=aH$ to the 
first-order term
of $\ln(aH/k)$. However, as shown in Fig.1 and Fig.2,
when $P_{S}(k_{COBE},a)$ starts to be approximately a constant,
$\ln(aH/k_{COBE})$ is larger than about 3. This raises another question about
the validity of the analytical expansion,
which can not be answered by an analytical argument. 
So a numerical study presented in this paper is necessary.

\thanks{
We thank R.H. Brandenberger and D. Lyth 
for reading the manuscript and comments. 
We also thank Y.S. Piao and Z.-H. Lin for discussions.
This work was supported in part by the National Natural Science Foundation of China.}

\newpage
\begin{figure}[htb]
\epsfxsize=6.0 in \epsfbox{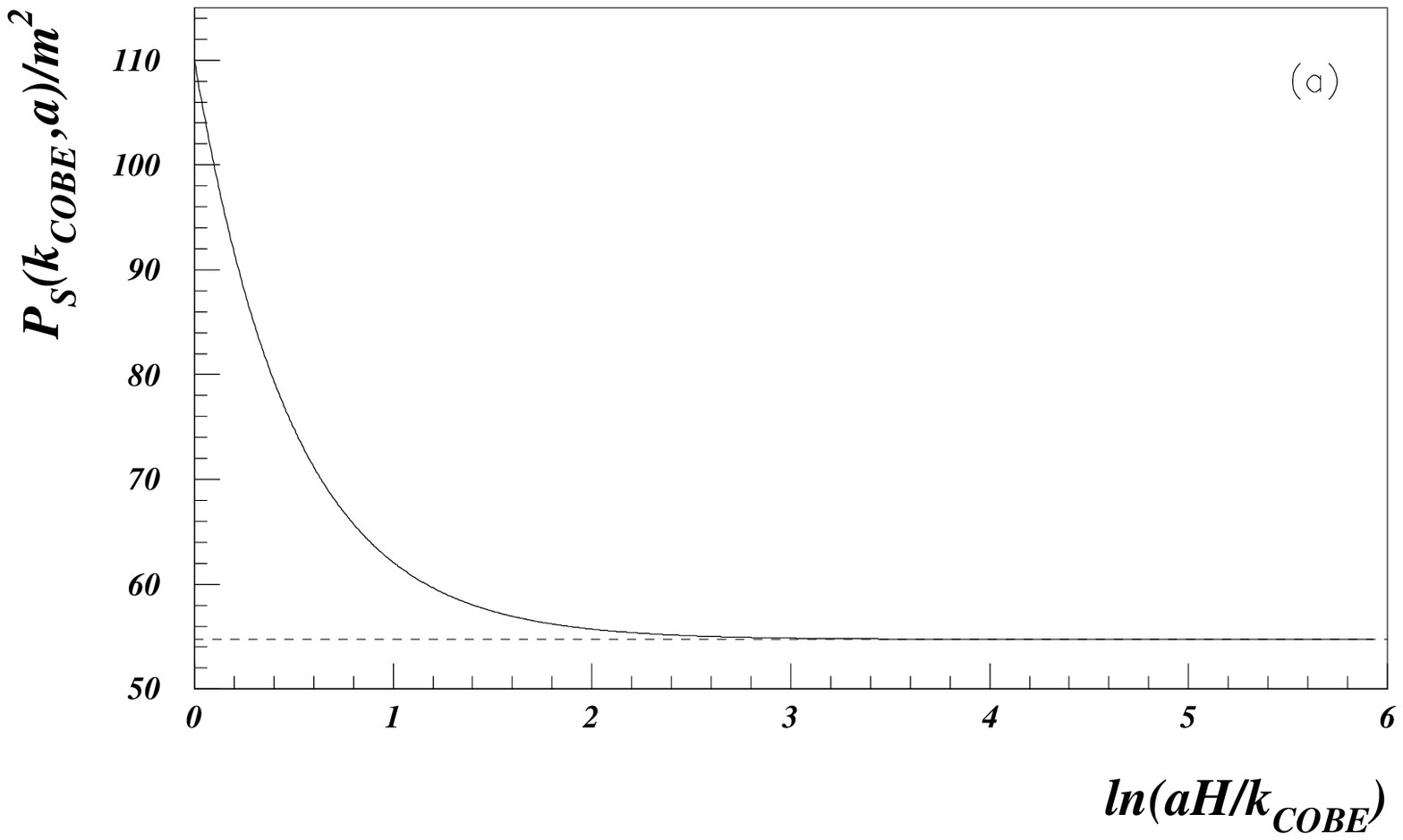}
\epsfxsize=6.0 in \epsfbox{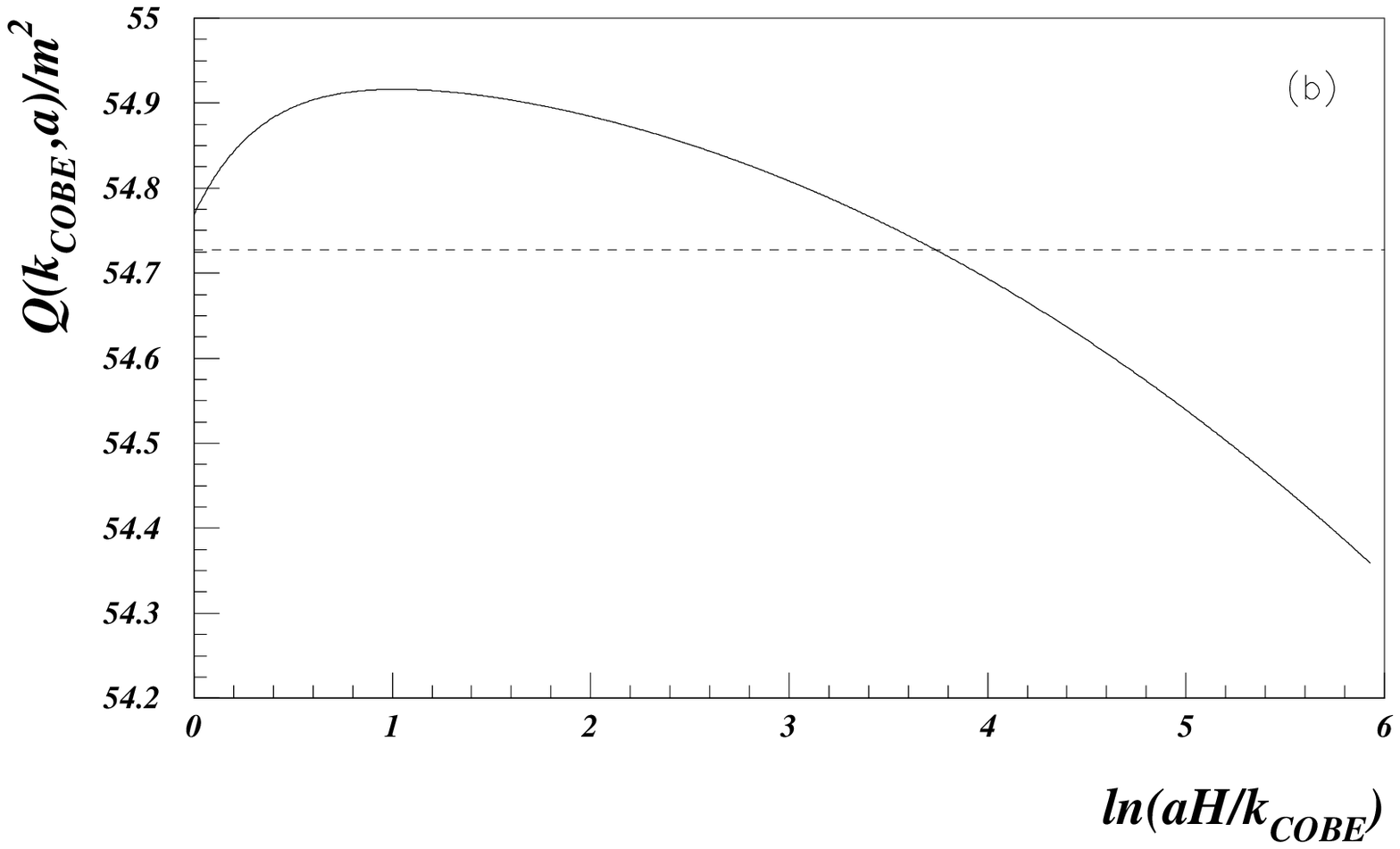}
\vskip 2mm
\caption{
Evolution of $P_{S}(k_{COBE},a)$ and $Q(k_{COBE},a)$ as a function of
$\ln (aH/k_{COBE})$
for model $V(\phi) =m^2\phi^2/2$
}
\end{figure}

\newpage

\begin{figure}
\epsfxsize=6.0 in \epsfbox{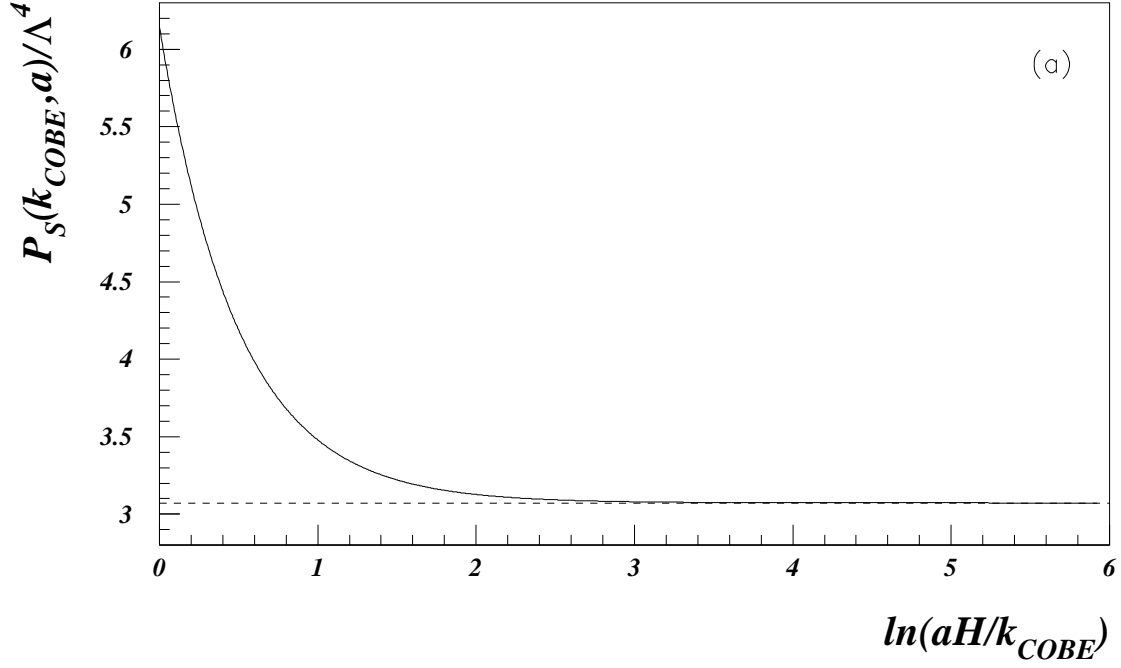}
\epsfxsize=6.0 in \epsfbox{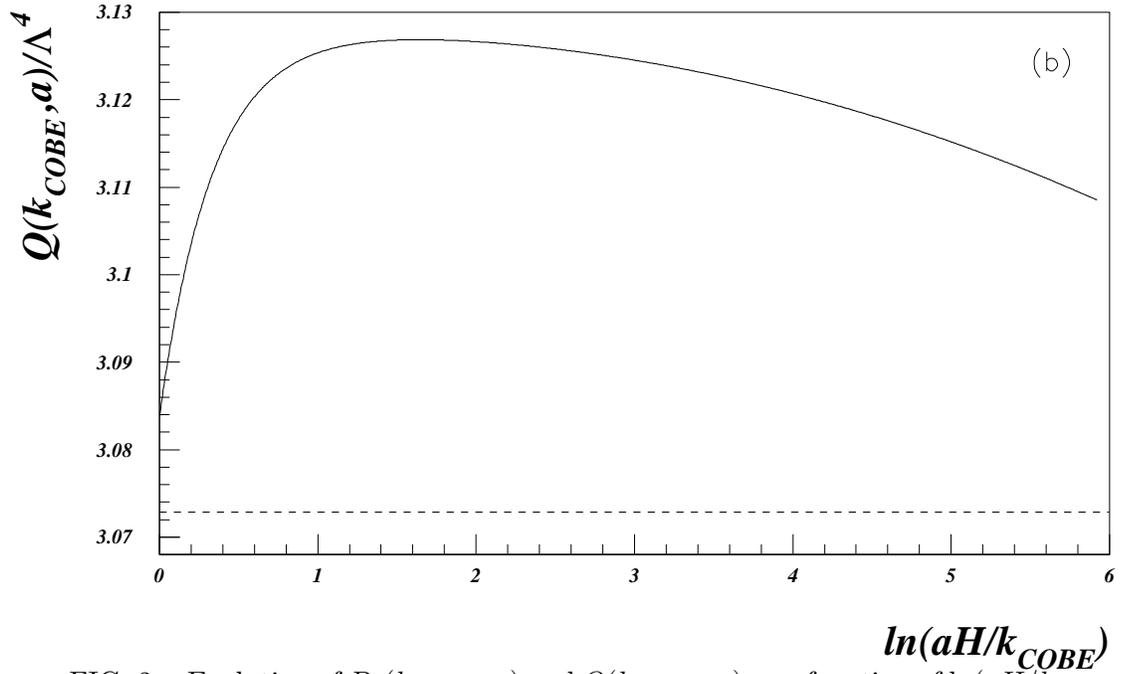}
\vskip 2mm
\caption{
Evolution of $P_{S}(k_{COBE},a)$ and $Q(k_{COBE},a)$ as a function of
$\ln (aH/k_{COBE})$
for model
$V=\Lambda^4[1+\cos(\phi/f)]$ with $f=1m_{pl}$.
}
\end{figure}

\begin{figure}
\epsfxsize=6.0 in \epsfbox{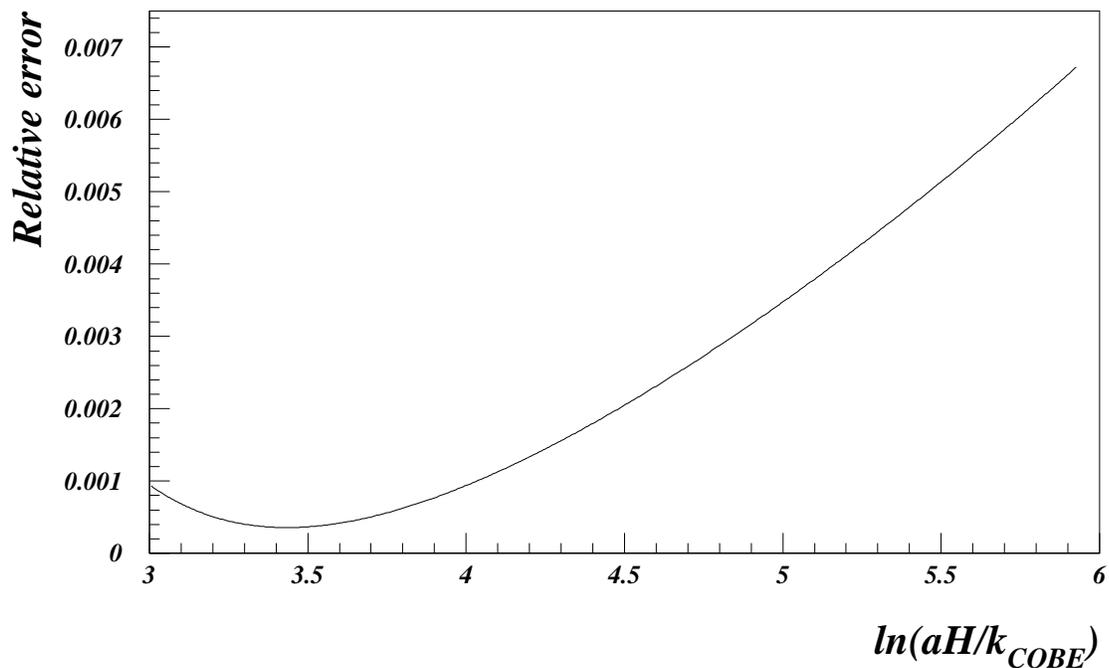}
\vskip 2mm
\caption{
The relative error between $Q(k_{COBE},a)$ and $P_{S}(k_{COBE},a)$
for model
$V(\phi) =m^2\phi^2/2$ ~.
}
\end{figure}

\begin{figure}
\epsfxsize=6.0 in \epsfbox{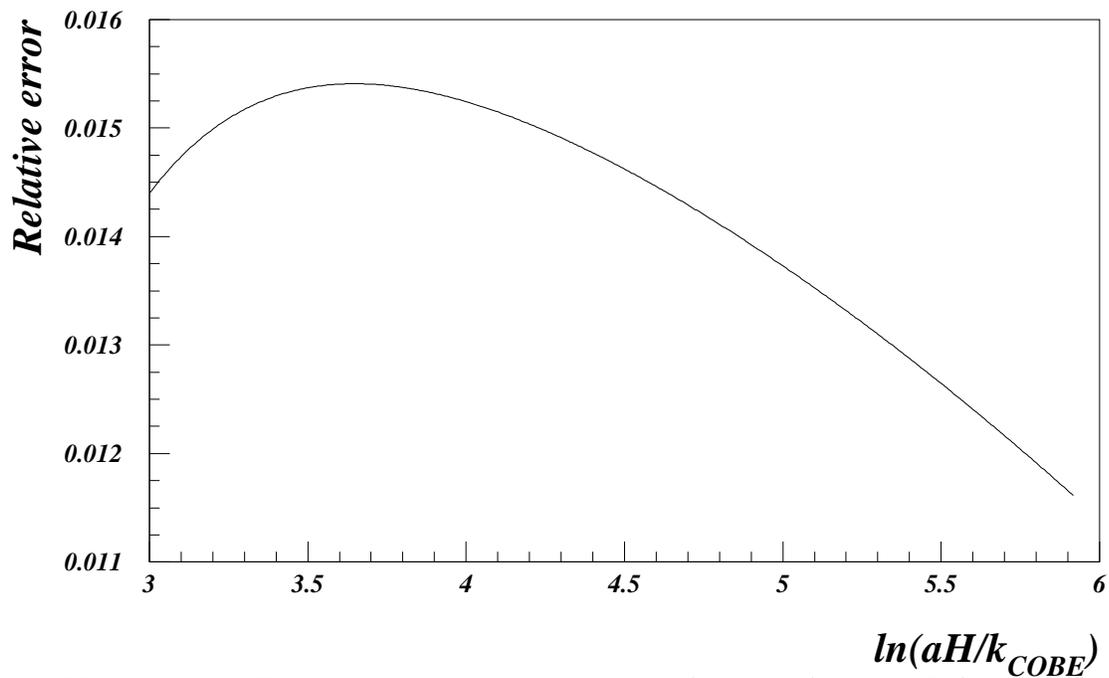}
\vskip 2mm
\caption{
The relative error between $Q(k_{COBE},a)$
and $P_{S}(k_{COBE},a)$
for model
$V=\Lambda^4[1+\cos(\phi/f)]$ with $f=1m_{pl}$~.
}
\end{figure}
\end{document}